\documentclass[12pt]{elsarticle} 

\vspace{5mm}

\usepackage{slashed}
\usepackage{graphicx}
\usepackage{amssymb}
\usepackage{mathtools}
\usepackage{bbold}
\usepackage{amssymb,latexsym}
\usepackage{amsmath,amsbsy,bbm}
\usepackage{multirow}
\usepackage[vcentermath]{youngtab}
\usepackage{nicefrac}
\usepackage[perpage]{footmisc}
\usepackage{wrapfig,lipsum,booktabs}
\usepackage{caption}
\usepackage{subcaption}
\usepackage{graphicx}
\usepackage{cjhebrew}

\usepackage{floatrow}
\usepackage[dvipsnames]{xcolor} 
\newfloatcommand{capbtabbox}{table}[][0.45\textwidth]

\newcommand{\lc}{\ensuremath{\Lambda_c}}
\newcommand{\fm}{\ensuremath{\,\text{fm}^{-1}}}
\newcommand{\abb}{\mbox{\ensuremath{A\oplus 1}}}

\newcommand{\lec}{C^\Lambda}
\newcommand{\led}{D^\Lambda}

\newcommand{\eg}{\textit{e.g.}~}
\newcommand{\ie}{\textit{i.e.}~}
\newcommand{\eftnopi}{\mbox{EFT$(\not \! \pi)$}}
\newcommand{\ve}[1]{\ensuremath{\boldsymbol{#1}}}

\newcommand{\be}{\begin{equation}}
\newcommand{\ee}{\end{equation}}

\newcommand{\Pe}{\text{\cjRL{p|}}}
\newcommand{\figref}[1]{Fig.~\ref{#1}}

\definecolor{blue}{HTML}{4169E1}
\definecolor{red}{HTML}{DC143C}
\definecolor{green}{HTML}{2E8B57}
\definecolor{mandarin}{HTML}{FF9933}

\usepackage[normalem]{ulem}

\makeatletter 
\def\ps@pprintTitle{%
 \let\@oddhead\@empty
 \let\@evenhead\@empty
 \def\@oddfoot{}%
 \let\@evenfoot\@oddfoot}
\makeatother

\begin{document}
\begin{frontmatter}

\title{Multi-fermion systems with contact theories}
\author{M.~Sch{\"a}fer}
\address{Nuclear Physics Institute of the Czech Academy of Sciences, 25069 \v{R}e\v{z}, Czech Republic}
\address{Czech Technical University in Prague, Faculty of Nuclear Sciences and Physical Engineering, B\v{r}ehov\'{a} 7, 11519 Prague 1, Czech Republic}
\author{L.~Contessi} 
\address{Racah Institute of Physics, The Hebrew University, 91904 Jerusalem, 
Israel} 
\address{ESNT, IRFU, CEA, Universit\'{e} Paris-Saclay, F-91191 Gif-sur-Yvette, France} 
\author{J. Kirscher}
\address{Theoretical Physics Division, School of Physics and Astronomy,
The University of Manchester, Manchester, M13 9PL, United Kingdom}
\author{J. Mare\v{s}}
\address{Nuclear Physics Institute of the Czech Academy of Sciences, 25069 \v{R}e\v{z}, Czech Republic}
\date{October 2018}

\begin{abstract}
We address the question of minimal requirements for the existence of quantum
 bound states. In particular, we demonstrate that a few-body system
with zero-range momentum-independent two-body interactions is unstable against
decay into clusters, if mixed-symmetry of its wave function is enforced.
We claim that any theory in which the two-body scattering length is much larger than any other scale involved exhibits such instability.
We exemplify this with
the inability of the leading-order pionless
effective field theory to describe stable states of $A>4$ nuclei.
A finite interaction range is identified as a sufficient
condition for a bound
mixed-symmetry system. The minimal value of this range depends on the
proximity of a system to unitarity, on the number of constituents, and
on the particular realization of discrete scale invariance of the three-body
spectrum.
\end{abstract}


\end{frontmatter}


\section{Introduction}

Systems that develop identical low-energy/long-range properties despite differences in 
their microscopic behavior are assigned to the same universality class. 
Critical exponents, the number, and nature
of few- and many-body states are examples of such common properties
which are related to a finite number of relevant interaction operators.
These operators define the minimal theory representing the universality class.
A class that is under intense study in a variety of fields comprises systems close to unitarity
namely those with a scattering length $a_0$
much larger than the interaction range $r_0$ and any other scattering parameter/scale.\footnote{
Notice the difference to a universality class that represents all systems
with a two-body
scattering volume $a_1$ which is, in addition to $a_0$, much larger than the other
scattering parameters.}
Systems belonging to this universality class are said to be scale invariant. They exhibit properties independent of the details of the interaction (see ~\cite{Braaten:2004rn,Naidon:2016dpf} for comprehensive reviews).
These features can be observed, for instance, in
%
%
%
%
hadronic molecules, nuclei, and atoms~(see, \eg,
~\cite{Tornqvist:1991ks,Voloshin:2003nt,Braaten:2003he,philli,tjon,PhysRevLett.81.69}) 
where very different degrees of freedom give rise to $r_0/a_0\to 0$\footnote{We adopt the canonical terms: {\it Resonant/Unitary} limit
if $|a_0|\to\infty$, and {\it zero-range/contact} limit if $r_0\to0$ with fixed $a_0$.}.
Notable examples of such universal behavior include the accumulation of an infinite number of
bound states at threshold in the three-boson system with a contact two-body interaction~\cite{Efimov:1971zz}~and the associated Thomas collapse~\cite{PhysRev.47.903}.
This collapse represents the emergence of a finite three-body scale $r_3\ll a_0$ in the absence of any finite two-body scale, with ensuing breaking of the continuous scale invariance down to its discrete version.
Particular realizations of the discrete scale invariance correspond to specific
physical systems, \eg, the triton~\cite{Bedaque:1998kg}.
The Tjon~\cite{tjon}~and Phillips~\cite{philli}~correlations, as well as spectra of multi-boson systems~\cite{PhysRevB.34.4571,blumig:2000,von_Stecher_2010,Gattobigio:2011ey,manybosons} are then consequences of the  universality extended beyond three particles and bound states. While the
Phillips correlation between three-body bound and scattering states was derived rigorously
from a resonant two-body interaction~\cite{Bedaque:1999ve}, numerical approaches were employed for
larger objects (see~\cite{Platter:2004he,Platter:2004zs} for Tjon's case).

All the above examples of shared properties pertain to states with total spatial symmetry. 
Very few instances of universal behavior are known for few- or many-body systems with a
singularly large two-body $S$-wave (unitary) scattering length which cannot reside in
pure $S$-wave states.
Two recent discoveries stand out: 
First, a zero-range theory was shown to stabilize~\footnote{We refer to stability of states with respect to a decay into any partition/set-of-clusters.}
systems of three 
fermions with two components of unequal mass only
if the mass ratio between the two species $\gtrapprox8.2$~\cite{Kartavtsev_2007}.
Second, a zero-range resonant two-body interaction was found to yield an unstable four-body
system of two-component fermions with equal mass as
inferred from the universal ratio between the scattering lengths of
two two-component dimers to the fermion-fermion
scattering length~\cite{petrov_dimerov, Petrov:2005zz,PhysRevA.92.053624}.
For nucleons, which are four-component fermions, non-$S$-wave features related to their bound, virtual and resonant states are mostly observed in systems with a larger number of particles.
Oxygen-16 was subsequently found to be unstable against four-$\alpha$ decay~\cite{Contessi:2017rww} with the contact-interaction condition realized within the framework of an effective field theory (EFT).
Adding a $P$-wave attraction nonperturbatively, however, yielded a stable ${}^{16}$O~\cite{Gattobigio:2019omi, PhysRevC.98.054301}.
The above studies hint towards a general characteristic of theories with $r_0/a_0\rightarrow 0$, namely,
the nonexistence of stable mixed-spatial-symmetry systems ($P$-wave systems, from hereon).

The focus of this article is on this hypothesis: Does
a unitary interaction preclude the existence of stable $P$-wave states in general, and if so,
what additional constraints have to be met to describe such states?
First, we complement the aforementioned EFT studies of few-nucleon systems with 6-, 7-, and 8-body calculations,
approaching the contact limit. 
Subsequently, we search for stable $P$-wave systems under the most favorable conditions:
Motivated by ~\cite{Kartavtsev_2007},~in which the binding of two noninteracting identical
fermions is induced by a third massive particle, we consider ($A+1$)-body systems where each includes a pair of identical fermions and $A$ particles forming a
massive spatially symmetric core. The motion of particles within this core is assumed
to be independent of the extra fermion and thus provides an average background field
that might give rise to an analogous binding effect.
As we will show, there is no evidence of $P$-wave binding even in this ideal scenario.
Thus we conclude that the absence of stable $P$-wave states in any system with a unitary
(or contact) $S$-wave interaction is one of its universal features.

This conjecture affects theoretical descriptions of all systems in this universality class,
\eg, the contact EFT for nuclei, which, as of now, cannot account for the
observed $P$-wave stability 
and might have to be augmented with an additional constraint.

\section*{Theory}
An effective field theory is the natural framework to study universality classes, as it contains the minimal number of operators to describe the low-energy features that define such classes.
The minimal EFT for nonrelativistic point particles forming shallow two-body and three-body $S$-wave states has been studied extensively (\eg, in ~\cite{Lepage:1997cs,vanKolck:1999mw, Bedaque:1998kg, Braaten:2004rn, Hammer:2017tjm, Hammer:2019poc}).
Its Hamiltonian formulation at leading order (LO) comprises zero-range two- and three-body vertices which depend on the renormalization parameter~$\Lambda$:
\begin{align}
H = - \sum_{i<j} \frac{\hbar^2}{2m}\ve{\nabla}_{ij}^2+ \lec \sum_{i<j}{\delta_\Lambda(\ve{r}_i-\ve{r}_j)} 
\,+\nonumber\\
\led \sum_{ i<j<k \atop \text{cyc} }\delta_\Lambda(\ve{r}_i-\ve{r}_j)\delta_\Lambda(\ve{r}_i-\ve{r}_k).
\label{eq:hamiltonian}
\end{align}
In the expansion of any resultant amplitude, the LO is given by all Born
terms depending solely on the coupling constants $\lec$ and $\led$. 
Parameters representing refinements
enter perturbatively at the order given na\"ively by their mass
dimension. 
In this work, a Gaussian regulator 
\mbox{$\delta_\Lambda(\ve{x}) \propto\Lambda^3 e^{-\frac{\Lambda^2}{4}\ve{x}^2}$} is used (such that $\Lambda\propto r_0^{-1}$).
The thus-induced dependence of $\lec$ and $\led$ on $\Lambda$ was calibrated to
the binding energy of a single state in the two- \mbox{($B(2)$)} and three-body
($B(3)$) system, respectively.
The few-body problem is thereby specified at LO with four parameters: The particle's mass
(here, $m=938~$MeV), the number of particles together with their statistics,
and the dimer and trimer binding energies.

We use this approach
to consider few-body nuclei, as well as systems with ($A+1$) equal-mass constituents, which contain only one pair of indistinguishable fermions. 
Therefore, a subset of $A$ distinguishable particles can be described by a symmetrical spatial wave function (\eg, the $S$-shell), as stipulated in the introduction.
These arrangements will be referred to as \abb~to highlight the core size $A$.
We approach the unitary limit
by increasing the ratio \mbox{$\Pe:=B(3)/B(2)$} and by taking $\Lambda \propto 1/r_0 \rightarrow +\infty$.
Exact unitarity is realized with $B(2)=0$ ($a_0\rightarrow +\infty$) in combination with $B(3)=1$~MeV; deviations from it with $B(2)=1$~MeV and $B(3)\in\lbrace3,\,4\rbrace$~MeV.
In the case of nuclear systems, we consider a spin and isospin-independent interaction. The parameters are then tuned to yield the deuteron and triton binding energies $B(2)=2.22$~MeV and $B(3)=8.48$~MeV, respectively. Such a parametrization is in accordance with the recent EFT formulation of $S$-shell nuclei in an expansion about the unitary limit, which is (iso)spin symmetric at LO \cite{Konig:2016utl} and thus referred to as SU(4)-symmetric pionless EFT (\eftnopi).

The necessary fits for $\led$ employ diagonalizations in bases built with the Stochastic
Variational Method (SVM,~\cite{Suzuki:1631377}) and the
Resonating Group Method (RGM,~\cite{PhysRev.52.1083,hmh-rrgm}). The coupling constant $\lec$ is
determined via a Numerov integration of the appropriate one-dimensional
radial Schr\"odinger equation.


\section*{Results:}

\begin{figure}[b!]
\centering
\includegraphics[width=0.8\linewidth]{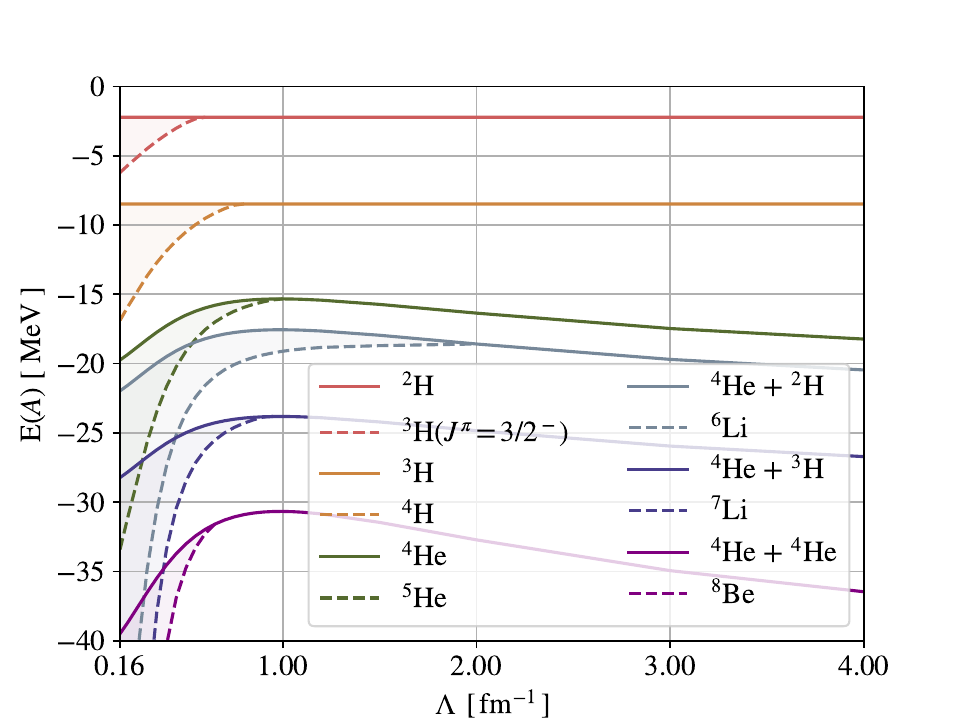} 
\caption{Cutoff dependence of nuclear ground-state energies
obtained with \eftnopi~at LO. For $A \leq 4$, solid lines represent nuclei with
spatially symmetric ground-state wave-function components. For $A>4$,
solid lines mark the lowest decay threshold into two spatially symmetric fragments.
The large-cutoff region $\Lambda > 4$~fm$^{-1}$ is not shown to improve readability. We calculated energies for cutoffs up to $\Lambda = 10$~fm$^{-1}$ and found results in agreement with the expected $E\propto\Lambda^{-1}$
behavior~\cite{Contessi:2017rww}
(see also~\figref{fig:threshold}, left panel).
}
\label{fig:nuclear}
\end{figure}

\begin{figure}
    \centering
        \centering
        \includegraphics[width=\linewidth]{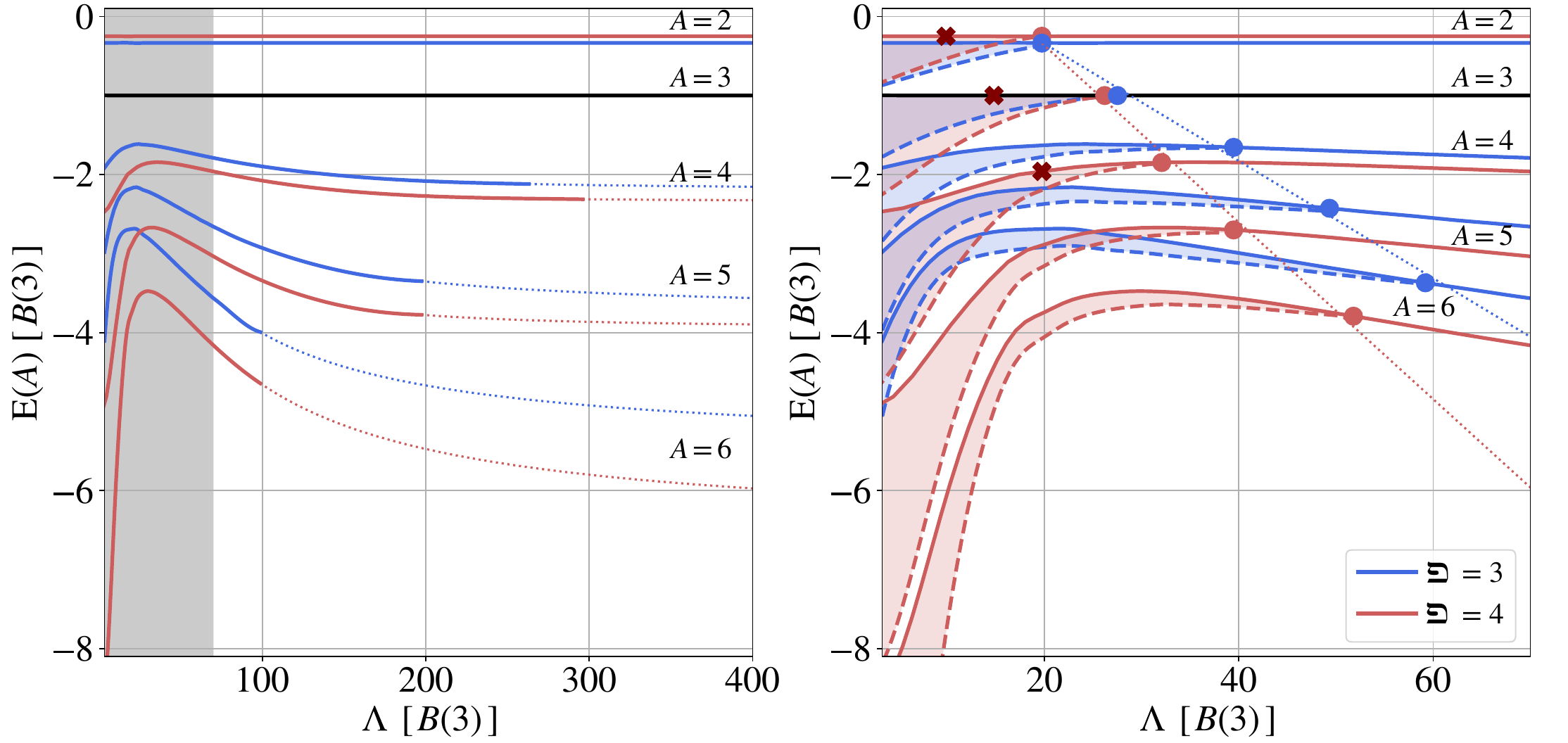} 
        \caption{Left panel: Cutoff dependence of ground-state energies of systems comprising $A$ distinguishable fermions described by the Hamiltonian in ~\eqref{eq:hamiltonian}~calibrated to $B(2)=1~$MeV and \mbox{$B(3)\in\lbrace3,4\rbrace~$MeV} (blue and red). Dotted lines denote extrapolation curves. Right panel: Small-cutoff region
        (gray area in the left panel) including ground-state energies of mixed-symmetry \abb~systems (dashed lines). The linear-in-$A$ trend of critical cutoffs $\lc$ (dots) is highlighted
        by dotted lines. Red crosses indicate critical cutoffs in case of a scattering volume equal to zero.}
        \label{fig:threshold}
\end{figure}

Our results, as obtained within the above defined theory
for the lowest energy eigenvalues of nuclear systems of up to eight nucleons, are shown
in~\figref{fig:nuclear}.
Particle-stable $P$-shell nuclei (dashed lines) are found only below a critical cutoff $\lc<2\fm$.
For larger cutoffs, \ie shorter-ranged interactions, they become unstable with respect to
fragmentation, \eg, \mbox{$^3$H($J^\pi=3/2^-$)$\to$ $^2{\rm H}+ n$}, $^4\text{H}\to\,^3\text{H} + n$,
$^5\text{He}\to\alpha +n$, $^6\text{Li}\to\alpha+{}^2\text{H}$, $^7\text{Li}\to\alpha + {}^3\text{H}$,
 and \mbox{$^8\text{Be}\to \alpha+\alpha$}.
The cutoff value below which a stable state exists is system dependent:
\begin{equation}
\label{eq:lc-order}
\lc(^3\text{H})<\lc(^8\text{Be})<\lc(^4\text{H})<\lc(^7\text{Li})<\lc(^5\text{He})<\lc(^6\text{Li})\;\;.
\end{equation}
Na\"ively, we expect three characteristic parameters of a system to control the value of $\lc$:
the total number of particles $A$, the typical size of the smaller of the two decay fragments and the number of its constituents.
With the size and the number of particles in the smaller fragment the attraction due to particle
exchange rises, increasing stability even for shorter interaction ranges.
The maximal $\lc$ in relation~\eqref{eq:lc-order}, which is due to the large
deuteron fragment in ${}^6$Li, demonstrates this effect.
Moreover, with the increase of the total number of particles the quantity of interacting pairs and triplets grows.
Keeping the smaller fragment size constant, we observe
that this effect is almost linear for a relatively small number of particles $A \leq 6$ (see the discussion of~\figref{fig:threshold}).
For all considered nuclei, we find $\lc$ to be of the order of the breakdown scale\footnote{We
adopt the canonical estimate of $m_\pi$ for the breakdown scale of \eftnopi~around which
substructure of the nucleons represented by the pion as the lightest meson becomes observable.}
of the \eftnopi.
The instability is therefore invariant with respect to renormalization-group (RG)
transformations with $\Lambda\gg m_\pi$.
On the one hand, 
this implies that no new constraints are needed to prevent the collapse of these larger
nuclei.
On the other hand, the three-parameter theory postdicts correctly only the experimentally
observed instability of 
$^3\text{H}(3/2^-),\,^3n,\,^4\text{H},\,^{3,4}\text{Li},\,\text{and}~^5\text{He}$.
The isotopes $^{6,7}\text{Li}$ and $^8$Be with $J^\pi=1^+$, $3/2^-$, and $0^+$, respectively, which are known to sustain stable
states\footnote{Without Coulomb repulsion, $^8$Be is found to be stable ~\cite{AFZAL:1969zz,Higa:2008dn}.}~are
not correctly reproduced in that respect.
Interestingly, once a stable $^8$Be ground state is found with $J^\pi=0^+$ at $\Lambda<\lc$ the theory orders excited states in the
rotational spectrum correctly, namely, $0^+$, $2^+$, and $4^+$. 
The latter two states emerge in a form of bound excited states for $\Lambda<0.4~{\rm fm^{-1}}$.

Having encountered $P$-shell instability in the nuclear case, we continue with an
investigation into
whether or not the same type of instability persists in a more general category
of systems.
In particular, we increase the number of distinguishable fermions $A$ forming the totally spatially symmetric core ($S$-shell). Starting with the $A$-body core, we consider different values of $\Pe$ (see chapter Theory) which parametrizes such systems arbitrarily close to unitarity. In the left panel of~\figref{fig:threshold}, we show ground-state energies $E(A)$, $2\leq A\leq 6$, as functions of the cutoff $\Lambda$ for $\Pe=3~(B(2)=1~{\rm MeV},~B(3)=3~{\rm MeV})$ and $\Pe=4~(B(2)=1~{\rm MeV},~B(3)=4~{\rm MeV})$. As $E(2)$ and $E(3)$ are used as renormalization constraints of the EFT, they are cutoff independent by construction. The ground-state energies of the 4-, 5-, and
6-body systems assume an asymptotic $\Lambda^{-1}$ dependence. We
visualized this latter behavior with a fit of $E(\Lambda)=\alpha+\beta/\Lambda$ to the corresponding $4\leq A\leq6$ ground state energies (dotted lines in the left panel of~\figref{fig:threshold}). The same behavior was reported in \cite{Betzalel:2016} for 4-, 5-, and 6-body Bose systems. The inequality \mbox{$B(A)/B(3)\Big\vert_{\Pe=3}<B(A)/B(3)\Big\vert_{\Pe=4}$} indicates that the universal ratios of the $A$- and 3-body binding energies are approached from below when taking the unitary limit. Furthermore, removing the finite two-body scale by recalibrating EFTs at unitarity ($\Pe\rightarrow \infty$), we obtain the ratio between binding energies of 4- and 3-body systems $B(4)/B(3)$ in agreement with ~\cite{Hammer:2006ct,2009NatPh...5..417V}.

Next, we assess the stability of \abb~systems including one additional fermion indistinguishable from one of the $A$ particles in the core.
In such systems, we do not find stable states with total orbital angular momentum $L_\text{total}=0$ for all considered cutoff values. For $L_\text{total}=1$,
in contrast, we do obtain stable states for small cutoff values (right panel of~ \figref{fig:threshold}). 
When we increase the cutoff towards the contact limit, all considered \abb~systems become unbound at some critical value $\lc$ (filled dots) with respect to a dissociation into
the $A$-body core and one free fermion. For fixed $\Pe$, these $\lc$'s increase approximately linearly with the system size: \mbox{$\lc\propto A \leq 6$} (dotted lines).

Results presented thus far, support the hypothesis of unstable $P$-wave states for $A\leq 6$, only. 
To investigate conceivable deviations from the linear dependence of $\lc$ for $A>6$,
particularly in form of divergences, we employ a two-fragment, local, single-channel,
resonating-group approximation (to be detailed in an upcoming communication as an extension of
~\cite{PhysRev.52.1083,Naidon_2016}). 
We thereby formulate the (\abb)-body system as a two-body problem of a
``frozen'' core with $A$ particles in $S$-shell and one particle which cannot occupy a state
in the same shell.
The accuracy of the approximation is expected to
improve at larger $A$ because of the increasing binding per particle in the $S$-wave cluster with its dimension~\cite{manybosons}. More specifically,
the gap between $B(A)$ and $B(A-1)$ widens, and
a core excitation by the out-of-shell particle is inhibited.
In practice, this approximation is employed for large cutoffs where the system is unstable and, therefore,
clustered into two fragments. Then the cutoff is decreased to the critical point $\lc$ at which
these two fragments just bind (for $\Lambda\ll\lc$, the approximation becomes less accurate).  
To parametrize the $A$-body fragment wave function we employ a one-parameter representation of the symmetric core as a product of harmonic oscillator ground states.
For $A \leq 7$, this parameter is fitted to the SVM results for the rms radius of the core.
For larger $A$, we match the core wave function with the liquid drop model formula $r_\text{rms}\propto A^{1/3}$ as 
it was found to be a good description for such systems in ~\cite{manybosons}.

\begin{figure}[t!]
\centering
\includegraphics[width=0.8\linewidth]{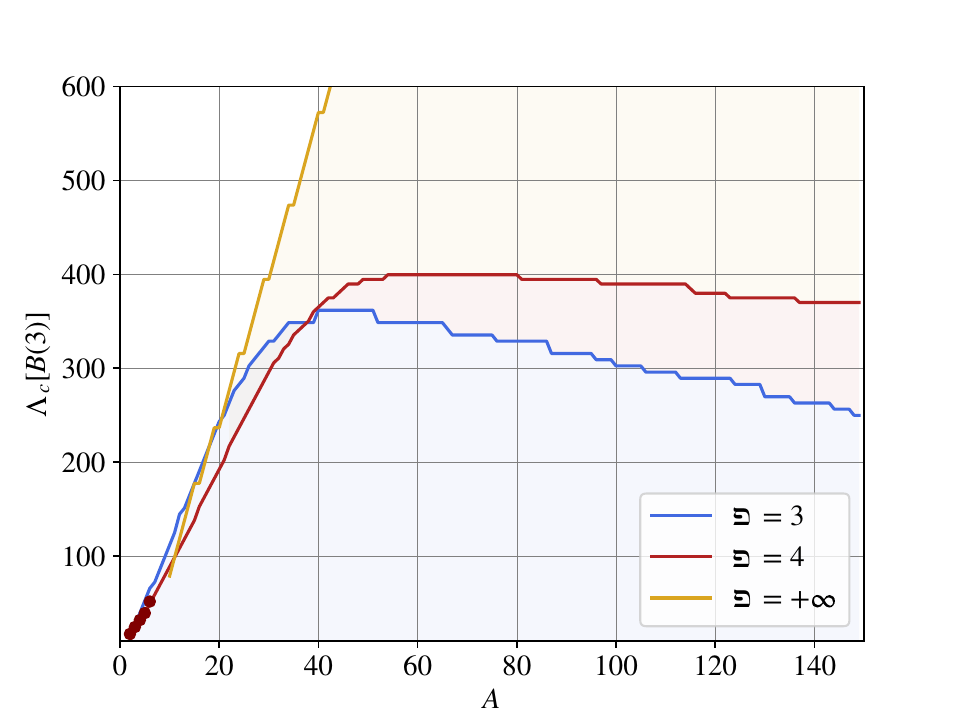} 
\caption{Dependence of the critical cutoff $\lc$ on the number of core particles
$A$. The SVM few-body results (red dots) are shown for $A\leq6$ and $B(3)=4~$MeV (see right panel of~\figref{fig:threshold})~along with single-channel resonating-group approximations for $A<150$ (lines). The unitary limit
(yellow) was realized with $B(2)=0$ ($a_0\rightarrow+\infty$) and $B(3)=1~$MeV; deviations from it
with $B(2)=1~$MeV and $B(3)\in\lbrace3,4\rbrace~$MeV (blue, red). 
In the shaded regions, the respective theories do sustain stable \abb~states,
while systems above the lines are unstable.
The step-like change in the curves results from a numerical criterion for the onset of binding and
can be removed systematically. }
\label{fig:RGM}
\end{figure}
Away from unitarity, we find $\lc(A)$ to abandon the linear increase found for small $A$.
$\lc$ increases up to a maximum number of particles $A^*$, while this maximum and the associated $\lc(A^*)$
both increase with $\Pe$ (compare maxima of the blue and red curves in~\figref{fig:RGM}). At unitarity,
we find a linear dependence up to $A > 100$ (yellow curve in~\figref{fig:RGM}).
The existence of $A^*$ and the developing nonlinear relation between $\Lambda_c$ and $A$ thus appear to be effects of the deviation from unitarity.
In combination with the microscopic results for lighter systems, this confirms the main hypothesis of this work, namely,
that a finite system at unitarity cannot
support stable states in the zero-range limit, i.e. $\Lambda \rightarrow \infty,$ if it contains indistinguishable fermions.

As far as nuclear physics, as a specific incarnation of this general class of systems, is concerned,
the description of atomic nuclei within the LO \eftnopi~will not yield stable states in systems heavier than $^4$He.
Consequently, such nuclei, as well as all other $P$-wave systems with stable states
belong to a different universality class which is defined by at least one new low-energy constraint.
A first step to identify it
is to analyse the mechanism behind the stability of the $P$-wave systems for $\Lambda<\lc$.
Of all artefacts introduced by the finite range of the regulated contact interaction,
a finite effective range in the two-body $S$-wave channel and a nonzero attractive two-body \mbox{$P$-wave} interaction are expected to dominate. 
Both contribute to the attraction in the \abb~system but their relative significance is obscure.
In other words, the finite-range interaction does not only describe a large \mbox{$S$-wave} scattering length but
also other finite parameters of the effective-range expansion as the effective range $r_0$ and the scattering volume $a_1$.
To shed light onto their relative importance, we project the two-body interaction into even partial waves in order to eliminate the effect of the scattering volume.
The resultant reduction of $\lc$ for all $A$ is significant compared with its
magnitude (red crosses in~\figref{fig:threshold}), which is now solely an effect of
even-partial-wave finite-range artefacts. 
Besides this decrease in $\lc$, the $P$-wave binding energy is also reduced by a substantial amount
as a consequence of the removal of the odd-partial-wave interactions.
We infer from these results a similar significance of the finite $r_0$ and $a_1$ for the stability of the nuclear $P$-wave systems.
Therefore, we cannot
conclusively identify an enhancement of either of them
as the perturbative deviation from unitarity needed to provide the binding.
We also cannot dismiss a conceivable fine tuning of the theory regarding one or both
of these parameters which
would define a new universality class that binds the systems.
%


\section*{Conclusion}
We studied numerically the stability of multi-fermion systems with contact theories. We represented these theories by the leading order of a nonrelativistic effective field theory whose renormalization conditions were chosen to be set by properties of spatially totally symmetric shallow dimer and trimer states, respectively.
In this framework, we performed few-body calculations up to $A=8$ using a
stochastic variational method with a correlated Gaussian basis. We found no $P$-shell nuclear states with less than 8 particles stable against fragmentation into spatially symmetric sub-components once the cutoff-renormalization scale of the theory exceeded a system-dependent critical value.
\addtolength{\textheight}{1.5cm}
\newpage
\addtolength{\topmargin}{-1.5cm}%
We found the same result for the stability of increasingly larger systems.
Starting with an $A\oplus1$-body structure -- \ie, $A$ distinguishable fermions form a totally spatially symmetric core, and the extra particle is a fermion with a distinguishable twin in the core -- we gradually increased the core size $A$.
These systems dissociate in the $A$-body core and the one free fermion once the cutoff is increased beyond an $A$-dependent critical value.
Using the two-fragment approximation of the resonating-group model, we assessed the dependence of the critical cutoff on the core size $2\leq A \leq 150$ and the proximity to unitarity. With a fixed and finite scattering length, the critical value increases with $A$ up to a maximum before decreasing from thereon. In the unitary case, it rises linearly up to at least $A\sim100$, suggesting that for any finite-range interaction a number of particles could be found which realize a stable $P$-wave composite. However, as the dependence on $A$ does not indicate a divergence of the critical cutoff for any of the analysed systems, we infer the absence of stable mixed-symmetry states in the contact limit.
Our calculations indicate that fermionic systems, whose dynamics is given by momentum-independent two- and three-body contact terms, are stable only if their spatial state is totally symmetric. States of mixed symmetry, as enforced by the presence of identical fermions, are predicted to be unstable. The appropriate EFT for the
description of stable mixed-symmetry states, larger nuclei in particular, thus defines a universality class that differs from unitary universality.

Finally, we assess the effect of two possible deviations from the unitary limit:
A nonzero two-body $S$-wave effective-range, and a nonzero two-body $P$-wave scattering volume. Both were found of similar importance for the binding of $P$-wave states, and our study is in this respect inconclusive as it cannot give preference for one over the other. In addition, it is not clear how these corrections should be treated -- as perturbative refinements of the unitary theory; or in a nonperturbative fashion, defining a new universality class. Three scenarios ought to be explored:

\begin{enumerate}
\item The contact effective field theory supports a shallow resonance pole at leading order which is
invariant with respect to renormalization-group transformations and 
can be mutated into a bound state with perturbative
insertions of sub-leading operators.
\item A resonance pole exists under identical conditions but requires a nonperturbative mechanism to be moved onto a bound-state pole.
\item The contact theory does not yield amplitudes for processes with states of mixed spatial symmetry and poles in its convergence radius. In order to be applicable
to observables affected by such poles,
the theory must create them nonperturbatively.
\end{enumerate}
To the best of our knowledge, it is unclear which scenario pertains to \mbox{$P$-wave} systems, \eg,~the shallow resonance found in the five-nucleon system. Our preliminary results for the $2\oplus1$ system corroborate the third scenario. However, scale invariance is broken for more than two particles, and the associated emergence of a finite scale may, in principle, pin the resonance also to a finite energy. The pole trajectory could thus be qualitatively different, and it remains to be calculated rigorously in order to advance the development of a contact effective field theory for $P$-wave systems. 

\paragraph*{Acknowledgments}
We thank N.~Barnea,  M.~Birse, U.~van Kolck, N.~Walet for insightful
discussions.
M.S. and J.M. were supported by the Czech Science Foundation GACR Grant No.19-19640S.
L.C. and J.K. acknowledge support from ``Espace de Structure et de r\'eactions
Nucl\'eaire Th\'eorique'' (ESNT, http://esnt.cea.fr) at CEA-Saclay, where this work
was partially carried out.
L.C. was also supported by the Pazy Foundation and by the
Israel Science Foundation Grant No. 1308/16.


\bibliographystyle{ieeetr}
\bibliography{Thebibliography.bib}
\end{document}